\documentclass[graybox,natbib]{svmult}
\usepackage{mathptmx}       % selects Times Roman as basic font
\usepackage{helvet}         % selects Helvetica as sans-serif font
\usepackage{courier}        % selects Courier as typewriter font
\usepackage{type1cm}        % activate if the above 3 fonts are
\usepackage{makeidx}         % allows index generation
\usepackage{graphicx}        % standard LaTeX graphics tool
\usepackage{multicol}        % used for the two-column index
\usepackage[bottom]{footmisc}% places footnotes at page bottom

\usepackage{amsmath, amssymb,bbm,color}

\newtheorem{prop}{Proposition}

\newtheorem{thm}{Theorem}

\newcommand{\mfrac}[2]{\genfrac{}{}{1.0pt}{}{#1}{#2}}

\newcommand{\f}{{\mathbf f}}

\newcommand{\card}{\mathop{\mathrm{card}}}
\newcommand{\supp}{\mathop{\rm{supp}}}

\newcommand{\ZZ}{{\mathbbm{Z}}}
\newcommand{\NN}{{\mathbbm{N}}}

\makeindex             % used for the subject index
\begin{document}

\title*{Orbits of Bernoulli Measures in Cellular Automata}
\author{Henryk Fuk\'s}
\institute{Henryk Fuk\'s \at Department of Mathematics and Statistics, Brock University,  St. Catharines,
ON, Canada \email{hfuks@brocku.ca}}
\maketitle

\section*{Glossary}
\begin{description}

\item[\textbf{Configuration space}] Set of all bisequences of symbols from the alphabet $\mathcal{A}$ of $N$ symbols,  $\mathcal{A}=\{0,1,\ldots, N-1\}$, denoted by $\mathcal{A}^\ZZ$.  Elements of $\mathcal{A}^\ZZ$ are called configurations and denoted by bold lowercase letters: $\mathbf{x}$, $\mathbf{y}$, etc.

\item[\textbf{Block} or \textbf{word}] A finite sequence of symbols of the alphabet $\mathcal{A}$. Set of all
blocks of length $n$ is denoted by $\mathcal{A}^n$, while the sent of all possible blocks of all lengths by 
$\mathcal{A}^\star$. Blocks are denoted by bold lowercase letters $\mathbf{a}$, $\mathbf{b}$, $\mathbf{c}$, etc.
Individual symbols of the block $\mathbf{b}$ are denoted by indexed italic form of the same letter, $\mathbf{b}=b_1,b_2,\ldots,b_n$.
To make formulae more compact, commas are sometimes dropped (if no confusion arises), and we simply write
$\mathbf{b}=b_1b_2 \ldots b_n$. 

\item[\textbf{Cylinder set}]  For a block $\mathbf{b}$ of length $n$, the cylinder set generated by $\mathbf{b}$ 
and anchored at $i$ is the subset of configurations such that symbols at positions from
$i$ to $i+n-1$ are fixed and equal to symbols in the block $\mathbf{b}$, while   the remaining symbols are arbitrary.
Denoted by $[\mathbf{b}]_i=\{ \mathbf{x}\in {\mathcal{A}}^\mathbb{Z}: \mathbf{x}_{[i,i+n)}=\mathbf{b} \}$.

 \item[\textbf{Cellular automaton}] In this article, cellular automaton is understood as  a map $F$
in the space of shift-invariant probability measures over the configuration space $\mathcal{A}^\ZZ$.  To define
$F$, two ingredients are needed, a positive integer $r$ called radius and a function
$w: \mathcal{A} \times \mathcal{A}^{2r+1} \to [0,1]$, whose values are called transition probabilities.
The image of a measure $\mu$ under the action of $F$ is then defined by probabilities of cylinder sets,
$(F\mu)([\mathbf{a}]_i)=\sum_{\mathbf{b}\in \mathcal{A}^{|\mathbf{a}|+2r}} w(\mathbf{a}| \mathbf{b}) \mu([\mathbf{b}]_{i-r})$
where  $i \in \mathbb{Z}$, $\mathbf{a} \in \mathcal{A}^{\star}$, and where $w(\mathbf{a}| \mathbf{b})$ is defined as
$ w(\mathbf{a}| \mathbf{b}) = \prod_{j=1}^{|\mathbf{a}|} w(a_j|b_{j}b_{j+1}\ldots b_{j+2r})$. Cellular automaton is called deterministic if transition probabilities take values exclusively in the set $\{0,1\}$, otherwise it is called
probabilistic.

\item[\textbf{Orbit of a measure}] For a given cellular automaton $F$ and a given shift invariant probability
measure $\mu$, the orbit of $\mu$ under $F$ is a sequence $\mu, F\mu, F^2 \mu,  F^3 \mu, \ldots$. The main subject of this article are orbits of Bernoulli measures on $\{0,1\}^\ZZ$, that is, measures parametrized by $p\in [0,1]$ and defined by
$\mu_p([\mathbf{b}])=p^{\#_1(\mathbf{b})} (1-p)^{\#_0(\mathbf{b})}$, where $\#_k(\mathbf{b})$ denotes number of
symbols $k$ in $\mathbf{b}$.

 \item[\textbf{Block probability}] Probability of occurence of a given block $\mathbf{b}$ (or word) of symbols. Formally
 defined as a measure of the cylinder set generated by the block $\mathbf{b}$ and anchored at $i$, and denoted by $P(\mathbf{b})=\mu([\mathbf{b}]_i)$. In this article we are exclusively dealing with shift-invariant probability
measures, thus $\mu([\mathbf{b}]_i)$ is independent of $i$.
Probability of occurence of a  block $\mathbf{b}$ after $n$ iterations
of cellular automaton $F$ starting from initial measure $\mu$ is
 denoted by $P_n(\mathbf{b})$ and defined as $P_n(\mathbf{b})=(F^n\mu)([\mathbf{b}]_i)$. Here again we assume shift 
invariance, thus $(F^n\mu)([\mathbf{b}]_i)$ is independent of $i$.

 \item[\textbf{Short/long block representation}] Shift invariant probability measures on $\mathcal{A}^\ZZ$ are 
unambiguously determined by block probabilities $P(\mathbf{b})$, $\mathbf{b} \in \mathcal{A}^\star$. For a given $k$,
probabilities of blocks of length $1,2, \ldots, k $ are not all independent, as they have to satisfy 
measure additivity conditions, known as Kolmogorov consistency conditions. One can show that only
$(N-1)N^{k-1}$ of them are linearly independent. If one declares as independent the set of
$(N-1)N^{k-1}$ blocks chosen so that they are as short blocks as possible, one can express the remaining blocks probabilities in terms of these. An algorithm
for selection of shortest possible blocks is called short block representation. If, on the other hand, one chooses the longest possible blocks to be declared independent, this is called long block representation.

 % \item[\textbf{Transition probabilities}] Ala ma kota
 % \item[\textbf{Bayesian approximation}] Ala ma kota
  \item[\textbf{Local structure approximation}] Approximation of points of the orbit of a measure $\mu$
   under a given cellular automaton $F$ by Markov measures, that is, measures maximizing entropy and
   completely determined by probabilities of blocks of length $k$. The number $k$ is called the order or level
   of local structure approximation. 

\item[\textbf{Block evolution operator}] 
When the cellular automaton rule of radius $r$  is deterministic, its transition probabilities take values
in the set $\{0,1\}$.  For such rules and for $\mathcal{A}=\{0,1\}$, define the local function $f:\mathcal{A}^{2r+1}
\to \mathcal{A}$ by
$f(x_1,x_2,\ldots x_{2r+1})$ $=w(1|x_1,x_2,\ldots x_{2r+1})$ for all $x_1,x_2,\ldots x_{2r+1} \in \mathcal{A}$.
A block evolution operator corresponding to $f$ is a mapping
 $\f:\mathcal{A}^{\star} \mapsto \mathcal{A}^{\star}$ defined for
  $a=a_0a_1 \ldots a_{n-1}\in \mathcal{A}^{n}$ by 
$f(a) = \{ f(a_i,a_{i+1},\ldots,a_{i+2r})\}_{i=0}^{n-2r-1}$. For a deterministic cellular automaton $F$
its local function is denoted by the corresponding lowercase italic form of the same letter, $f$, while
the block evolution operator is the bold form of the same letter, $\mathbf{f}$. The set of preimages
of the block $\mathbf{a}$ under $\f$ is called block preimage set, denoted by $\f^{-1}(\mathbf{a})$.

\item[\textbf{Complete set set}]  A set of words $C=\{\mathbf{a}_1,\mathbf{a}_2,\mathbf{a}_3,\ldots\}$  is called complete 
with respect to a CA rule $F$ if for every  $\mathbf{a}\in C$ and $n\in\mathbb{N}$,  $P_{n+1}(\mathbf{a})$ can be expressed
as a linear combination of $P_n(\mathbf{a}_1),P_n(\mathbf{a}_2),P_n(\mathbf{a}_3),\ldots$.
\end{description}

\tableofcontents

\section{Introduction}
In both theory and applications of cellular automata (CA), one of the most natural and most frequently
encountered problems  is what one could call the \emph{density response problem}:
If the proportion of ones (or other symbols) in the initial configuration drawn from a Bernoulli distribution is known, what is the expected proportion of ones (or other symbols) after $n$ iterations of the CA rule?
One could rephrase it in a slightly different form: if the probability of occurence of a given symbol in 
an initial configuration is known, what is the probability of occurrence of this symbol after $n$
iterations of this rule?

A similar question could be asked about the probability of occurence of longer blocks of symbols
after $n$ iterations of the rule. Due to complexity of CA dynamics, there is no hope to answer questions like this in a general form, for an arbitrary rule. The situation is somewhat similar to hat we encounter in the theory
of  differential equations: there is no general algorithm for solving initial value problem for an arbitrary rule,
but one can either  solve it approximately (by numerical method), or, for \emph{some} differential equations, one
can construct the solution formula in terms of elementary functions.

In cellular automata,  there are also two ways to make  progress. One is to use some approximation techniques, and compute approximate values
of the desired probabilities. Another is to focus on narrower classes of CA rules, with sufficiently simple dynamics,
and attempt to compute these probabilities in a rigorous ways. Both these approaches are discussed in this article.

We will treat cellular automata  as maps in the space of Borel 
shift-invariant probability measures,  equipped
with the so-called weak$\star$ topology \citep{KurkaMaas2000,Kurka2005,Pivato2009,FormentiKurka2009}.
In this setting, the aforementioned problem of computing block probabilities can be
posed as the  problem of  determining the orbit of  given initial measure $\mu$ (usually a Bernoulli measure) under the action of a given cellular automaton. Since computing the complete orbit of a measure is, in general, very difficult, approximate methods have been developed.
The simplest of these methods is called the mean-field theory, and has its origins in statistical physics \citep{Wolfram83}.
The main idea behind the mean-field theory is to approximate the consecutive iterations of 
the initial measure by  Bernoulli measures, ignoring correlations between sites.  While this approximation is obviously very crude, it is sometimes quite useful in applications. 

In 1987,  H. A. Gutowitz, J. D. Victor, and B. W. Knight proposed a generalization
of the mean-field theory for cellular automata which, unlike the mean-field theory,  takes  into account 
correlations between sites, although only in an approximate way  \citep{gutowitz87a}. The basic idea is to approximate 
the consecutive iterations of the initial measure by Markov measures, also called finite block measures.
Finite block measures of order $k$ are completely determined by probabilities of blocks of length $k$.
For this reason,  one can construct a map on these block probabilities, which, when iterated, approximates probabilities of occurrence of the same  blocks in the actual orbit of a given  cellular automaton. The construction of Markov measures is  based on  the idea of  ``Bayesian extension'', introduced in 1970s and
80s in the context of lattice gases
\citep{Brascamp71,Fannes84}. The local structure theory produces remarkably good approximations of probabilities
of small blocks, provided that one uses sufficiently high order of the Markov measure.

For deterministic CA, if one wants to compute probabilities of small block exactly, without using any approximations, 
one has to study combinatorial structure of preimages of these block under the action of the rule.
In many cases, this reveals some regularities which can be exploited in computation of block probabilities.
For a number of elementary CA rules, this approach has been used to construct probabilities of short blocks,
typically block of up to three symbols. For probabilistic cellular automata, one can try to compute $n$-step transition probabilities,
and in some cases these transition probabilities are expressible in terms of elementary functions. This allows
to construct formulae for block probabilities.

In the rest of this article we will discuss how to construct shift-invariant probability measures over the space of 
bisequences of symbols, and
how to describe such measures in terms of block probabilities. We will then define cellular automata as
maps in the space of measures and discuss orbits of shift-invariant probability measures under these maps. Subsequently, the local structure approximation
will be discussed as a method to approximate orbits of Bernoulli measures under the action of cellular automata.
The final sections presents some known examples of cellular automata, both deterministic and probabilistic,  for which  elements of the orbit
of the Bernoulli measure (probabilities of short blocks) can be determined exactly.
\section{Construction of a probability measure}
Let ${\mathcal{A}}=\{0,1,\ldots,N-1\}$ be called an \emph{alphabet}\index{alphabet}, or a \emph{symbol set}\index{symbol set}, 
and let $X={\mathcal{A}}^\mathbb{Z}$ be called the \emph{configurations space}.
The \emph{Cantor metric}\index{Cantor metric} on $X$ is defined as $d(\mathbf{x},\mathbf{y})=2^{-k}$, where $k=\mathrm{min} \{ |i|: \mathbf{x}_i
\neq  \mathbf{y}_i\}$. $X$ with the metric $d$ is a Cantor space, that is, compact, totally
disconnected and perfect metric space.
A finite sequence of elements of ${\mathcal{A}}$, $\mathbf{b}=b_1b_2\ldots, b_{n}$ will be called a \emph{block}\index{block of symbols}  (or \emph{word})
 of length $n$.
Set of all blocks of elements of ${\mathcal{A}}$ of all possible lengths will be denoted by ${\mathcal{A}}^{\star}$.
\emph{A cylinder set}\index{cylinder set} generated by the block $\mathbf{b}=b_1b_2\ldots, b_{n}$ and anchored at $i$  is defined as
\begin{equation}
[\mathbf{b}]_i=\{ \mathbf{x}\in {\mathcal{A}}^\mathbb{Z}: \mathbf{x}_{[i,i+n)}=\mathbf{b} \}.
%\{ \mathbf{x}\in {\mathcal{A}}^\mathbb{Z}:
%x_i=b_1, x_{i+1}=b_2, \ldots, x_{i+n-1}=b_n\},
\end{equation}
When  one of the indices $i,i+1,\ldots, i+n-1$ is equal to zero, the cylinder set will be called \emph{elementary}.
The collection (class) of all elementary cylinder sets of $X$ together with the
empty set and the whole space $X$ will be denoted by $\mathit{Cyl}(X)$. 
One can show that  $\mathit{Cyl}(X)$ constitutes a semi-algebra over $X$. 
% This, in addition to aforementioned   closure under 
% the intersection, means that the set difference of two elementary cylinder sets is a finite union of elementary
% cylinder sets.
% 
% We will now introduce the notion of a measure on the semi-algebra of cylinder sets.
% Let $\mathcal{D}$ be a semi-algebra. A map $\mu: {\mathcal{D}} \to [0,\infty]$ is called a \emph{measure} on $\mathcal{D}$ if 
% it is countably additive and $\mu(\varnothing)=0$. By \emph{countable additivity} we mean that for any sequence
% $\{A_i\}_{i=1}^{\infty}$ of pairwise disjoint sets belonging to  $\mathcal{D}$ such that $\bigcup_{i=1}^{\infty} A_i \in {\mathcal{D}}$,
% \begin{equation}
% \mu\left(\bigcup_{i=1}^{\infty}A_i\right) = \sum_{i=1}^{\infty} \mu(A_i).
% \end{equation}
%For measures on the semi-algebra of cylinder sets, countable additivity is implied by finite additivity.
Moreover, one can  show that any finitely additive map $\mu: \mathit{Cyl}(X) \to [0,\infty]$  for which  $\mu(\varnothing)=0$ 
is a measure on the semi-algebra of elementary cylinder sets $\mathit{Cyl}(X)$.

The semi-algebra of elementary cylinder sets\index{semi-algebra of elementary cylinder sets}, equipped with a measure
 is still  ``too small'' a class of subsets of $X$ to support all requirements
of probability theory. For this we need a $\sigma$-algebra\index{$\sigma$-algebra}, that is, a class
of subsets of $X$ that is closed under the complement and under the countable unions of its members.  
Such $\sigma$-algebra can be defined as an ``extension'' of $\mathit{Cyl}(X)$. The smallest 
$\sigma$-algebra containing $\mathit{Cyl}(X)$ will be called \emph{$\sigma$-algebra generated
by $\mathit{Cyl}(X)$}. As it turns out, it is possible to extend a measure on semi-algebra to
the $\sigma$-algebra generated by it, by using Hahn-Kolmogorov theorem\index{Hahn-Kolmogorov theorem}.

In what follows, we will only consider measures for which $\mu(X)=1$ (probability measures). 
Moreover, we will only limit our attention to the case
of translationally invariant measures (also called shift-invariant),
by requiring that, for all $\mathbf{b} \in {\mathcal{A}}^{\star}$,  $\mu([\mathbf{b}]_i)$ is 
independent of $i$. To simplify notation, we then define $P: {\mathcal{A}}^{\star} \to [0,1]$ as
\begin{equation}
 P(\mathbf{b}):=\mu([\mathbf{b}]_i).
\end{equation}
Values $P(\mathbf{b})$ will be called \emph{block probabilities}\index{block probabilities}. 
Application of Hahn-Kolmogorov theorem to the case of shift-invariant probability measure $\mu$ yields the following result.
\begin{thm}\label{extensionfromblocks}
 Let $P: {\mathcal{A}}^{\star} \to [0,1]$  satisfy
the conditions
\begin{align} 
P(\mathbf{b})&= \sum_{a \in {\mathcal{A}}} P(\mathbf{b}a) =\sum_{a \in \cal{G}} P(a\mathbf{b})
\,\,\,\,\,\,\,\forall {\mathbf{b} \in {\mathcal{A}}^{\star}} ,\label{cons1}\\
1&=\sum_{a \in \cal{A}} P(a).\label{cons2}
\end{align}
Then $P$ uniquely determines shift-invariant probability measure\index{shift-invariant probability measure} on the $\sigma$-algebra
generated by elementary cylinder sets of $X$.
\end{thm}
The set of shift-invariant probability measures on the $\sigma$-algebra
generated by elementary cylinder sets of $X$ will be denoted by $\mathfrak{M}(X)$.

Conditions (\ref{cons1}) and (\ref{cons2}) are often called \emph{consistency conditions}\index{consistency conditions},
although they are essentially equivalent to measure additivity
conditions. Some consequences of consistency conditions in the context of cellular automata have been studied in detail by \cite{mcintosh2009one}. 
\section{Description of probability measures by block probabilities}
Since the probabilities $P(\mathbf{b})$ uniquely determine the probability measure, we can define a shift-invariant probability measure by specifying $P(\mathbf{b})$ for all $\mathbf{b}\in\mathcal{A}^\star$. Obviously, because of consistency conditions, block probabilities are not independent, thus in order to define the probability measure, we actually need to specify only \emph{some} of them, but not necessarily \emph{all} - the missing ones can be computed by using consistency conditions. 

Define $\mathbf{P}^{(k)}$ to be the column vector of all 
probabilities of blocks of length $k$ arranged in lexical order. For
example, for ${\mathcal{A}}=\{0,1\}$, these are
\begin{align*}
 \mathbf{P}^{(1)}&=[P(0), P(1)]^T,\\
\mathbf{P}^{(2)}&=[P(00),P(01),P(10),P(11)]^T,\\
\mathbf{P}^{(3)}&=[P(000),P(001),P(010),P(011),P(100),P(101),P(110),P(111)]^T,\\
&\cdots .
\end{align*}
The following result \citep{paper50} is a direct consequence of  consistency conditions of eq. (\ref{cons1}) and (\ref{cons2}).
\begin{prop}\label{propnrind}
Among all block probabilities constituting components of
 $\mathbf{P}^{(1)}$, $\mathbf{P}^{(2)}$,  $\dots, \mathbf{P}^{(k)} $
only $(N-1)N^{k-1}$ are linearly independent.
\end{prop}
 For
example, for $N=2$ and $k=3$, among $\mathbf{P}^{(1)} , \mathbf{P}^{(2)}, \mathbf{P}^{(3)}$ (which have,
in total, $8+4+2=14$ components), there are
only 4 independent blocks. These four block can be selected somewhat arbitrarily (but not completely
arbitrarily).  Two methods or algorithms for selection of independent blocks are especially convenient.

The first one is called \emph{long block representation}\index{long block representation}.
It is based on the following property (cf. ibid.).
\begin{prop} \label{hilowdependenceprop}
 Let $\mathbf{P}^{(k)}$ be partitioned into two sub-vectors, $\mathbf{P}^{(k)}=
(\mathbf{P}^{(k)}_{Top}$, $\mathbf{P}^{(k)}_{Bot})$, where $\mathbf{P}^{(k)}_{Top}$
contains first $N^k-N^{k-1}$ entries of $\mathbf{P}^{(k)}$, and
$\mathbf{P}^{(k)}_{Bot}$ the remaining $N^{k-1}$ entries. Then
\begin{equation}\label{depbyindep}
 \mathbf{P}^{(k)}_{Bot}=
\left[ \begin {array}{c} 0\\ \vdots \\  0
\\ 1\end {array} \right]
 - \left(\mathbf{B}^{(k)}\right)^{-1} \mathbf{A}^{(k)}\mathbf{P}^{(k)}_{Top}.
\end{equation}
\end{prop}
In the above, matrix $\mathbf{B}^{(k)}$  is constructed from zero $N^{k-1} \times N^{k-1}$ matrix  by
placing $-1$'s on the diagonal, and then filling the last row with  1's, so that
\begin{equation}
\mathbf{B}^{(k)}=\left[
 \begin {array}{rrrr} 
 -1&0&\cdots&0\\ 
 0&-1& & \\
 \vdots & &\ddots& \\
 1&1&1&1
 \end{array} \right].
\end{equation}
 The  matrix $\mathbf{A}^{(k)}$ 
is a bit more complicated,
\begin{equation}
 \mathbf{A}^{(k)} = [\mathbf{J}_{1} \mathbf{J}_{2} \ldots \mathbf{J}_{N-1}]+
[\,\underbrace{\mathbf{B}^{(k)}\,\,\, 
\mathbf{B}^{(k)} \ldots \mathbf{B}^{(k)}}_{N-1}\,],
\end{equation}
where $\mathbf{J}_{m}$ is an  $N^{k-1}\times N^{k-1}$ matrix in which $m$-th
row consist of all 1's, and all other entries are equal to 0.

The above proposition means that among block probabilities constituting components of
 $\mathbf{P}^{(1)}$ , $\mathbf{P}^{(2)} , \dots, \mathbf{P}^{(k)} $,
we can treat first $N^{k}-N^{k-1}$ entries of $\mathbf{P}^{(k)}$ as independent variables.
Remaining components of $\mathbf{P}^{(k)}$ can be obtained by using eq. (\ref{depbyindep}),
while  $\mathbf{P}^{(1)} , \mathbf{P}^{(2)} , \dots, \mathbf{P}^{(k-1)}$ can be obtained by
eq. (\ref{cons1}).

When applied to the  $N=2$ and $k=3$ case,
it yields the following choice of independent blocks: $P(000),P(001),P(010)$ and $P(011)$. The remaining 10 probabilities
can then be expressed as follows,
\begin{align} \label{longblockexamplek3}
 \left[ \begin {array}{c} P(100)\\
 P(101)\\ P(110)\\ 
P(111)\end {array} \right] &= \left[ \begin {array}{c} P(001)\\ 
-P(001)+P(010)+P(011)\\
 P(011)\\
1 -P(000)- P(001)-2\,P(010)-3\,P(011)\end {array} \right],  \nonumber \\
%%%%%%%%%%
\left[ \begin {array}{c} P(00)\\ P(01)\\ P(10)\\P(11)
\end {array} \right] &= \left[ \begin {array}{c} 
P(000)+P(001)\\ 
P(010)+P(011)\\ 
P(010)+P(011)\\ 
1-P(000)-P(001)-2\,P(010)-2\,P(011)
\end {array} \right],  \nonumber \\
%%%%%%%%%%%
 \left[ \begin {array}{c} P(0)\\ P(1)
\end {array} \right] &= \left[ \begin {array}{c}
 P(000)+P(001)+P(010)+P(011)\\
 1-P(000)-P(001)-P(010)-P(011)\end {array}
 \right]. 
\end{align}

Of course, this is not the only choice. Alternatively, we can choose as independent blocks
the shortest possible blocks. The algorithm resulting in such a choice  will be called
\emph{short block representation}\index{short block representation}. 
In order to describe it in a formal way, let us define a vector of
\emph{admissible entries} for short block representation, $\mathbf{P}^{(k)}_{adm}$, as follows.
Let us take vector $\mathbf{P}^{(k)}$ in which block probabilities are arranged in lexicographical order,
indexed by an index $i$ which runs from 1 to $N^k$. Vector $\mathbf{P}^{(k)}_{adm}$  consists of
all entries of $\mathbf{P}^{(k)}$ for which the index $i$ is not divisible by $N$ and for which
$i<N^k-N^{k-1}$. For example, for $N=3$ and $k=2$ we have
\begin{equation*}\mathbf{P}^{(2)}=
[P(00),P(01),P(02),P(10),P(11),P(12),P(20),P(21),P(22)]^T,
\end{equation*}
and we need to select entries with $i$ not divisible by 3 and $i<6$, which leaves $i=1,2,4,5$, hence
$$\mathbf{P}^{(2)}_{adm}=[P(00),P(01),P(10),P(11)]^T.$$

Vector of independent block probabilities in short block representation is now defined as
\begin{equation}
 \mathbf{P}^{(k)}_{short}
 =\left[ \begin{array}{c} \mathbf{P}^{(1)}_{adm}\\[0.2em] \mathbf{P}^{(2)}_{adm} \\ \vdots \\  \mathbf{P}^{(k)}_{adm}
\end{array} \right].
\end{equation}
The following result can be established.
\begin{prop} \label{shortblockprop}
Among  block probabilities constituting components of
 $\mathbf{P}^{(1)}$, $\mathbf{P}^{(2)},\dots$ $\mathbf{P}^{(k)} $,
we can treat entries of $\mathbf{P}^{(k)}_{short}$ as independent variables.
One can express all other components of $\mathbf{P}^{(1)}$, $\mathbf{P}^{(2)} , \dots, \mathbf{P}^{(k)} $
in terms of components $\mathbf{P}^{(k)}_{short}$.
\end{prop}
The exact formulae expressing  components of $\mathbf{P}^{(1)}$, $\mathbf{P}^{(2)} , \dots, \mathbf{P}^{(k)} $
in terms of components $\mathbf{P}^{(k)}_{short}$ are rather complicated, and can be found in \citep{paper50}.
As an example, for $N=2$ and $k=3$, this algorithm yields
$P(0)$, $P(00)$, $P(000)$ and $P(010)$ to be the independent block probabilities, that is, the components of
$\mathbf{P}^{(3)}_{short}$.  
The remaining 10 dependent blocks probabilities can be
expressed in terms of  $P(0)$, $P(00)$, $P(000)$ and $P(010)$.
\begin{align} \label{shortform3}
%%%%%%%%%%%%%%%%%%%%%%%%%%
\left[ \begin {array}{c} 
P(001)\\
P(011)\\  
P(100) \\
P(101)\\
P(110)\\
P(111)
\end {array} \right] &= 
\left[ \begin {array}{c} 
P(00)-P(000) \\ 
P(0)-P(00) -P(010) \\  
P(00)-P(000) \\  
P(0)-2 P(00)+P(000) \\ 
P(0)-P(00)-P(010)\\  
 1-3 P(0) +2 P(00) +P(010) 
\end {array} \right]. \nonumber \\
%%%%%%%%%%%%%%%%%%%%%%%%
%%%%%%%%%%%%%%%%%%%%%%%%%
\left[ \begin {array}{c} 
P(01)\\
P(10)\\  
P(11)
\end {array} \right] &= 
\left[ \begin {array}{c} 
P(0) -P(00) \\
P(0) -P(00) \\ 
 1-2 P(0)+P(00)
\end {array} \right], \nonumber  	\\
%%%%%%%%%%%%%%%%%%%%%%%
%%%%%%%%%%%%%%%%%%%%%%%%%
 P(1) &= 1-P(0). 
%%%%%%%%%%%%%%%%%%%%%%%%
\end{align}

\section{Cellular automata}

Let $w: \mathcal{A} \times \mathcal{A}^{2r+1} \to [0,1]$, whose values are denoted by $w(a|\mathbf{b})$
for $a \in \mathcal{A}$, $\mathbf{b} \in \mathcal{A}^{2r+1}$, satisfying
$\sum_{a \in \mathcal{A}} w(a|\mathbf{b})=1$, be called \emph{local transition function}\index{local transition function}
of \emph{radius} $r$, and its values  called \emph{local transition probabilities}\index{transition probabilities}.
\emph{Probabilistic cellular automaton}\index{Probabilistic cellular automaton}  with local 
transition function $w$ is a map $F: \mathfrak{M}(X) \to \mathfrak{M}(X)$ defined as
\begin{equation} \label{rulefed}
(F\mu)([\mathbf{a}]_i)=\sum_{\mathbf{b}\in \mathcal{A}^{|\mathbf{a}|+2r}} w(\mathbf{a}| \mathbf{b}) \mu([\mathbf{b}]_{i-r})
\mathrm{\,\, for\,\, all\,\,}  i \in \mathbb{Z}, \mathbf{a} \in \mathcal{A}^{\star},
\end{equation}
where we define
\begin{equation} \label{defw}
 w(\mathbf{a}| \mathbf{b}) = \prod_{j=1}^{|\mathbf{a}|} w(a_j|b_{j}b_{j+1}\ldots b_{j+2r}).
\end{equation}
When the function $w$ takes values in the set $\{0,1\}$, the corresponding cellular automaton is called 
\emph{deterministic cellular automaton}\index{deterministic cellular automaton}. 

For any shift-invariant probability measure $\mu \in \mathfrak{M}(X)$, we define the orbit of $\mu$ under $F$ as
\begin{equation}
 \{  F^n \mu \}_{n=0}^{\infty},
\end{equation}
where  $F^0\mu=\mu$.
Points of the orbit of $\mu$ under $F$ are uniquely determined by probabilities of cylinder sets. Thus,  
if we define, for $n \geq 0$, $P_n(\mathbf{a})=(F^n \mu)([\mathbf{a}]_i)$, then, for $\mathbf{a}\in \mathcal{A}^k$, eq. (\ref{rulefed}) becomes
\begin{equation} \label{recP}
 P_{n+1}(\mathbf{a})= \sum_{b \in \mathcal{A}^{|\mathbf{a}|+2r}} w(\mathbf{a}| \mathbf{b}) P_n(\mathbf{b}).
\end{equation}
 In the above we assume that $P_0(\mathbf{a})=\mu(\mathbf[a]_i)$.

Given the measure $\mu$, equations (\ref{recP}) define a system of recurrence relationship for block probabilities. Solving this
recurrence system, that is, finding $P_n(\mathbf{a})$ for all $n\in \NN$ and all $\mathbf{a}\in \mathcal{A}^{\star}$,
would be equivalent to determining the orbit of $\mu$ under $F$.
However, it is very difficult to solve these equations explicitly, and no general method for doing this is known.
To see the source of the difficulty, let us take ${\mathcal{A}}=\{0,1\}$ and let us consider the example of rule 14, 
for which local transitions probabilities are given by
\begin{align}
w(1|000) = 0, \,  w(1|001) = 1, \,  w(1|010) = 1, \,  w(1|011) = 1,  \nonumber \\
w(1|100) = 0, \,  w(1|101) = 0, \,  w(1|110) = 0, \,  w(1|111) = 0,
\end{align}
and $w(0|x_1x_2x_3)=1-w(1|x_1x_2x_3)$ for all $x_1,x_2,x_3 \in \{0,1\}$.
For $k=2$, eq. (\ref{recP}) becomes 
\begin{align} \label{r14exact}
 P_{n+1}(00)&=
P_n(0000)+
P_n(1000)+
P_n(1100)+
P_n(1101)+
P_n(1110)\nonumber \\
+&P_n(1111), \nonumber \\
P_{n+1}(01)&=
P_n(0001)+
P_n(1001)+
P_n(1010)+
P_n(1011),\nonumber \\
P_{n+1}(10)&=
P_n(0100)+
P_n(0101)+
P_n(0110)+
P_n(0111),\nonumber \\
P_{n+1}(11)&=
P_n(0010)+
P_n(0011).
\end{align}
It is obvious that this system of equations cannot be iterated over $n$, because on the left hand side
we have probabilities of blocks of length 2, and on the right hand side -- probabilities of blocks of
length 4. Of course, not all these probabilities are independent, thus it will be better to 
rewrite the above using short form representation. Since among block probabilities of length 2
only 2 are independent, we can take only two of the above equations, and express all block
probabilities occurring in them by their short form  representation, using eq. (\ref{shortform3}).
This reduces eq. (\ref{r14exact}) to
\begin{align} \label{r14exactreduced}
 P_{n+1}(0)&=1-P_n(0)+P_n(000), \nonumber \\
 P_{n+1}(00)&= 1-2 P_n(0)+P_n(00) + P_n(000).
\end{align}
Although much simpler, the above system of equations still cannot be iterated, because on the right hand
side we have an extra variable $P_n(000)$. To put it differently, one cannot reduce iterations
of $F$  to iterations of a finite-dimensional map (in this case, two-dimensional map). For this reason,
a special method  has been developed to approximate orbits of $F$ by orbits of finite-dimensional maps.

\section{Bayesian approximation}
For a given measure $\mu$,  it is clear that the knowledge of $\mathbf{P}^{(k)}$ is
enough to determine all $\mathbf{P}^{(i)}$ with $i<k$, by using consistency conditions. What about $i>k$? Obviously, since
the number of independent components in $\mathbf{P}^{(i)}$ is greater than in $\mathbf{P}^{(k)}$
for $i>k$, there is no hope to determine $\mathbf{P}^{(i)}$ using only $\mathbf{P}^{(k)}$.
It is possible, however, to approximate longer block probabilities by shorter block
probabilities using the idea of Bayesian extension\index{Bayesian extension}.

Suppose now that we want to approximate $P(a_1a_2\ldots a_{k+1})$ by 
$P(a_1a_2$ $\ldots a_{k})$. One can say that by knowing $P(a_1a_2\ldots a_{k})$ we know how
values of individual symbols in a block are correlated providing that symbols are not farther apart
than $k-1$. We do not know, however, anything about correlations on the larger length scale.
The only thing we can do in this situation is to simply neglect these higher length correlations,
and assume that if a block of length $k$ is extended by adding another symbol to it on the right, then
the the conditional probability of finding a particular value of that symbol does not significantly
depend on the left-most symbol, i.e.,
\begin{equation}
\frac{P(a_1a_2\ldots a_{k+1})}{P(a_1\ldots a_{k})} \approx \frac{ P(a_{2}\ldots a_{k+1})  }
{P(a_2 \ldots a_{k})}.
\end{equation}
This produces the desired approximation of $k+1$ block probabilities by $k$-block and $k-1$ block probabilities,
\begin{equation} \label{bayapprox}
 P(a_1a_2\ldots a_{k+1}) \approx \frac{ P(a_1\ldots a_{k}) P(a_{2}\ldots a_{k+1}) }
{P(a_2 \ldots a_{k})},
\end{equation}
where we assume that the denominator is positive. If the denominator is zero, then
we take  $P(a_1a_2\ldots a_{k+1}) =0$.
In order to avoid writing separate cases for denominator equal to zero, we define  ``thick bar'' fraction\index{``thick bar'' fraction} as
\begin{equation} \label{convention1}
 \mfrac{a}{b}:=\begin{cases}
               {\displaystyle \frac{a}{b}} & \mathrm{if\,\,} b \neq 0\\[1em]
                   0 & \mathrm{if\,\,} b = 0.      
                     \end{cases}
\end{equation}

Now, let $\mu \in \mathfrak{M}(X)$ be a measure with associated block probabilities $P: {\mathcal{A}}^{\star} \to [0,1]$,
$P(\mathbf{b})=\mu([\mathbf{b}]_i)$ for all $i\in \mathbb{Z}$ and $\mathbf{b}\in {\mathcal{A}}^{\star}$.
For $k>0$,  define
$\widetilde{P}: {\mathcal{A}}^{\star} \to [0,1]$ such that 
\begin{equation}
 \widetilde{P}(a_1a_2 \ldots a_{p})=
\begin{cases}
\,\,\,\,\,\,\,P(a_1a_2 \ldots a_{p}) & \mathrm{if\,\,} p \leq k ,\\[0.5em]
\displaystyle  \mfrac
{ 
\prod_{i=1}^{p-k+1} P(a_i \ldots a_{i+k-1})  
}
{ 
\prod_{i=1}^{p-k}  P(a_{i+1} \ldots a_{i+k-1}) 
}  & \mathrm{otherwise}.        
\end{cases}
\end{equation}
Then $\widetilde{P}$ determines a shift-invariant probability measure $\widetilde{\mu}^{(k)} \in \mathfrak{M}(X)$,
 to
be called \emph{Bayesian approximation of $\mu$ of order $k$}\index{Bayesian approximation}. 

When there exists $k$ such that Bayesian approximation of $\mu$ of order $k$ is equal to $\mu$, we call $\mu$
a \emph{Markov measure}\index{Markov measure} or a \emph{finite block measure}\index{finite block measure} of order $k$. The space of shift-invariant probability Markov measures of order $k$
will be denoted by $\mathfrak{M}^{(k)}(X)$,
\begin{equation}
 \mathfrak{M}^{(k)}(X)=\{\mu \in \mathfrak{M}(X): \mu=\widetilde{\mu}^{(k)}\}.
\end{equation}

It is often said that the Bayesian approximation ``maximizes entropy''. Le us define  \emph{entropy density}\index{entropy density} of shift-invariant measure  $\mu \in \mathfrak{M}(X) $ as
\begin{equation}
h(\mu)= \lim_{n \to \infty} - \frac{1}{n} \sum_{\mathbf{b} \in {\mathcal{A}}^n} P(\mathbf{b}) \log P(\mathbf{b}),
\end{equation}
where, as usual, $P(\mathbf{b})=\mu([\mathbf{b}]_i)$ for all $i\in \mathbb{Z}$ and $\mathbf{b}\in {\mathcal{A}}^{\star}$.
It has been established by  \cite{Fannes84} that
for any  $\mu \in \mathfrak{M}(X) $, the entropy density of the $k$-th order Bayesian approximation of $\mu$
is given by
\begin{equation}
h( \widetilde{\mu}^{(k)})=
\sum_{\mathbf{a} \in {\mathcal{A}}^{k-1}} P(\mathbf{a}) \log P(\mathbf{a})
-\sum_{\mathbf{a} \in {\mathcal{A}}^k} P(\mathbf{a}) \log P(\mathbf{a}),
\end{equation}
and that for any  $\mu \in \mathfrak{M}(X) $ and any $k>0$, the entropy density of $\mu$ does not exceed the
entropy density of its $k$-th order Bayesian approximation,
\begin{equation} \label{entropyineq}
 h(\mu) \leq h(\widetilde{\mu}^{(k)}).
\end{equation}
Moreover, one can show that the sequence of $k$-th order Bayesian approximations of
$\mu \in \mathfrak{M}(X)$ weakly converges to  $\mu$ as $k \to \infty$ \citep{gutowitz87a}.

Using the notion of Bayesian extension, H. Gutowitz \textit{et. al.} developed a method 
of approximating orbits of $F$, known as the \emph{local structure theory}\index{local structure theory} \citep{gutowitz87a,gutowitz87b}. 
Following these authors, let us define the \emph{scramble operator}\index{scramble operator} of order $k$, denoted by $\Xi^{(k)}$, and defined as
\begin{equation}
\Xi^{(k)} \mu=\widetilde{\mu}^{(k)}.
\end{equation}
The scramble operator, when applied to a shift invariant measure $\mu$, produces a Markov measure of order $k$ which agrees
with $\mu$ on all blocks of length up to $k$. The idea of local structure approximation is that each time step,
instead of just applying $F$, we apply scramble operator, then $F$, and then the scramble operator again.
This yields a sequence of Markov measures  $\nu_{n}^{(k)}$ defined recursively as
\begin{equation} \label{defrecnu}
  \nu_{n+1}^{(k)}= \Xi^{(k)} F \Xi^{(k)} \nu_{n}^{(k)},\,\,\,\,\, \nu_0^{(k)}=\mu.
\end{equation} 
The sequence defined as 
\begin{equation} \label{approxorbit}
\left\{ \left(\Xi^{(k)} F \Xi^{(k)} \right)^n \mu \right\}_{n=0}^{\infty}
\end{equation}
will be called the \emph{local structure approximation}\index{local structure approximation} of level $k$ of the exact orbit $\{  F^n \mu \}_{n=0}^{\infty}$.
Note that all terms of this sequence are Markov measures, thus the entire local structure approximation
of the orbit lies in $\mathfrak{M}^{(k)}(X)$. The following theorem describes the local structure approximation in a formal
way.
\begin{thm}
For any positive integer $n$, and for any shift invariant probability measure $\mu$, 
$\nu_n^{(k)}$ weakly converges to  $F^n \mu$ as $k \to \infty$.
\end{thm}

\section{Local structure maps}
A nice feature of Markov measures is that they can be entirely described by specifying probabilities
of a finite number of blocks. This makes construction of finite-dimensional maps
generating approximate orbits of measures in CA  possible.

Define  $Q_n(\mathbf{b})= \nu_{n}^{(k)}([\mathbf{b}])$.
Using definitions of $F$ and $\Xi$, eq. (\ref{defrecnu}) yields, for any  $\mathbf{a} \in \mathcal{A}^k$, 
\begin{equation} \label{lstmapcomponents}
  Q_{n+1}(\mathbf{a})= 
\sum_{\mathbf{a}\in \mathcal{A}^{|\mathbf{b}|+2r}} w(\mathbf{a}| \mathbf{b})  
\mfrac
{ 
\prod_{i=1}^{2r+1} Q_n ( \mathbf{b}_{[ i, i+k-1]})  
}
{ 
\prod_{i=1}^{2r} \sum_{c \in \mathcal{A}} Q_n (c \mathbf{b}_{[i+1,i+k-1]}) 
}.
\end{equation}

If we arrange $Q_n(\mathbf{a})$  for all $\mathbf{a} \in \mathcal{A}^k$ in lexicographical order
  to form a vector $\mathbf{Q}_n$, we will obtain
\begin{equation}
 \mathbf{Q}_{n+1} = L^{(k)} \left(\mathbf{Q}_{n}\right),
\end{equation} 
where $L^{(k)}: [0,1]^{|\mathcal{A}|^k} \to [0,1]^{|\mathcal{A}|^k}$ has components defined by eq. (\ref{lstmapcomponents}).
$L^{(k)}$  will be called
 \emph{local structure map}\index{local structure map} of level $k$. 

Of course, not all components of  $Q$ are independent, due to consistency conditions. We can, therefore,
further reduce dimensionality of local structure map to $(N-1)N^{k-1}$ dimensions. This will be illustrated 
for rule 14 considered earlier.

Recall that for rule 14, if we start with an initial measure $\mu$ and define $P_n(b)=(F^n \mu)[\mathbf{b}]$, then
\begin{align} 
 P_{n+1}(0)&=1-P_n(0)+P_n(000), \nonumber \\
 P_{n+1}(00)&= 1-2 P_n(0)+P_n(00) + P_n(000).
\end{align}
The corresponding local structure map can be obtained from the above by simply replacing $P$ by $Q$
and using the fact that block probabilities $Q$ represent Markov measure of order $k$, thus
\begin{equation}
 Q_n(000) = \mfrac{Q_n(00) Q_n(00)}{Q_n(0)}.
\end{equation}
Equations (\ref{r14exactreduced})  would then become
\begin{align}
 Q_{n+1}(0)&=1-Q_n(0)+\mfrac{Q_n(00)^2}{Q_n(0)}, \nonumber \\
 Q_{n+1}(00)&= 1-2 Q_n(0)+Q_n(00) +\mfrac{Q_n(00)^2}{Q_n(0)},
\end{align}
where $Q_0(0)=P_0(0)$,  $Q_0(00)=P_0(00)$.
The above is a formula for recursive iteration of a two-dimensional map, thus one could compute
$Q_{n}(0)$ and $Q_{n}(00)$ for consecutive $n=1,2 \ldots$ without referring to any other block
probabilities, in stark contrast with eq. (\ref{r14exactreduced}). Block probabilities $Q$
approximate exact block probabilities $P$, and the quality of this approximation varies
depending on the rule. Nevertheless, as the order of approximation $k$ increases, values
of $Q$ become closer and closer to $P$, due to the weak convergence of $\nu_n^{(k)}$ to $F^n \mu$.

As an illustration of this convergence, let us consider a probabilistic rule defined by
\begin{align}\label{rule18asynch}
w(1|000) = 0, \,  w(1|001) = \alpha, \,  w(1|010) = 1-\alpha, \,  w(1|011) = 1-\alpha,  \nonumber \\
w(1|100) = \alpha, \,  w(1|101) = 0, \,  w(1|110) = 1-\alpha, \,  w(1|111) = 1-\alpha,
\end{align}
and $w(0|x_1x_2x_3)=1-w(1|x_1x_2x_3)$ for all $x_1,x_2,x_3 \in \{0,1\}$, where $\alpha \in [0,1]$ is
a parameter. This rule is known as $\alpha$-asynchronous elementary rule 18 \cite{fatjca09}, because
for $\alpha=1$ it reduces to elementary CA rule 18. It is known that for this rule, if one starts with
initial symmetric Bernoulli measure $\mu_{1/2}$, then $\lim_{n \to \infty} P_n(1)=0$
if $\alpha \leq \alpha_c$, and $\lim_{n \to \infty} P_n(1)>0$ if $\alpha> \alpha_c$,
where $\alpha_c\approx 0.7$. This phenomenon can be observed in simulations, if one iterates 
the rule for large number of time steps $T$ and records $P_T(1)$.
The graph of $P_T(1)$ as a function of $\alpha$ for $T=10^4$, obtained
by such direct simulations of the rule, is shown in Figure 1. To approximate $P_T(1)$ by local structure
theory, one can construct
local structure map of order $k$ for this rule, iterate it $T$ times,
and obtain $Q_{T}(1)$, which should approximate $P_T(1)$. The graphs of
$Q_{T}(1)$ vs. $\alpha$, obtained this way, are shown in Figure 1 as dashed lines. One can clearly see that
as $k$ increases, the dashed curves approximate the graph of $P_{T}(1)$ better and better.
%%%%%%%%%%%%%%%%%%%%%%%
\begin{figure}
\includegraphics[width=10cm]{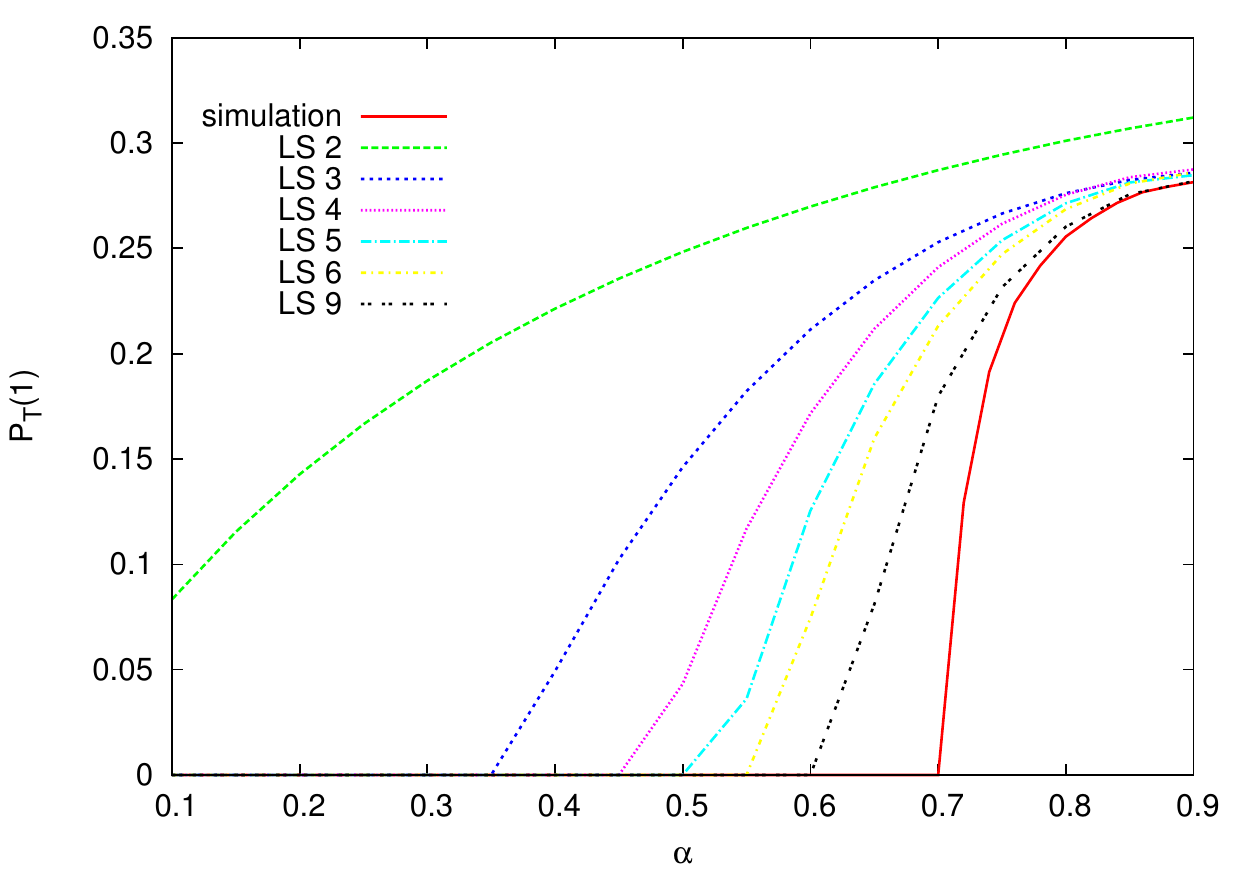}
\caption{Graph of $P_{T}(1)$ for $T=10^4$ as a function $\alpha$ for probabilistic CA rule defined in eq. (\ref{rule18asynch}).
Continuous line represents values of $P_{T}(1)$ obtained by Monte Carlo simulations, and dashed lines
values of $Q_{T}(1)$ obtained by iterating local structure maps of level $k=2,3,4,5$ and $9$.}
\end{figure}

For some simple CA rules, the local structure approximation is exact. Such is the case of idempotent rules\index{idempotent rule},
that is, CA rules for which $F^2=F$. \cite{gutowitz87a} found that this is also the case for 
what he calls linear rules, toggle rules\index{toggle rule}, and asymptotically trivial rules\index{asymptotically trivial rule}.

\section{Exact calculations of probabilities of short blocks along the orbit}

If approximations provided by the local structure theory are not enough, one can attempt to
compute orbits of Bernoulli measures exactly. Typically, it is not possible to obtain
expressions for all block probabilities $P_n(\mathbf{a})$ along the orbit, yet one can often compute 
$P_n(\mathbf{a})$ if $\mathbf{a}$ is short, for example, containing just one, two, or three symbols.

For elementary CA rules, the behaviour of $P_n(1)$ as a function of $n$ has been studied
extensively by many authors, starting from  \cite{Wolfram83}, who determined numerical values of $P_\infty(1)$ for 
a wide class of CA rules and postulated exact values for some of them. Later one exact values
of  $P_n(1)$ have been established for some elementary rules, and in some cases,  $P_n(\mathbf{a})$
has been computed for all $|\mathbf{a}|\leq 3$. We will discuss these results in what follows.

When the rule is deterministic, transition probabilities in eq. (\ref{rulefed}) take values
in the set $\{0,1\}$. Let us consider \emph{elementary cellular automata}\index{elementary cellular automata}, that is, binary rules for which $N=1$, ${\mathcal{A}}=\{0,1\}$ and the radius $r=1$. For such rules, define the \emph{local function}\index{local function} $f$ by
$f(x_1,x_2,x_3)=w(1|x_1x_2x_3)$ for all $x_1,x_2,x_3 \in \{0,1\}$. Elementary CA with the local
local function $f$ are usually identified
by their Wolfram number $W(f)$, defined as \citep{Wolfram83}
\begin{equation*} \label{code3}
W(f)=\sum_{x_1,x_2,x_3=0}^{1}f(x_1,x_2,x_3)2^{(2^2x_1+2^1x_2+2^0x_3)}.
\end{equation*}

A \emph{block evolution operator}\index{block evolution operator} corresponding to $f$ is a mapping
 $\f:\mathcal{A}^{\star} \mapsto \mathcal{A}^{\star}$ defined as follows. 
Let  $a=a_0a_1 \ldots a_{n-1}\in \mathcal{A}^{n}$
where $n \geq 3 $. Then 
\begin{equation}
\f(a) = \{ f(a_i,a_{i+1},a_{i+2})\}_{i=0}^{n-3}.
\end{equation}
For elementary CA eq. (\ref{rulefed}) reduces to
\begin{equation} 
(F\mu)([\mathbf{a}])=\sum_{\mathbf{b}\in \f^{-1}(\mathbf{a})}  \mu([\mathbf{b}]),
\end{equation}
where we dropped indices indicating where the cylinder set is anchored (we assume shift-invariance of measure $\mu$), 
and where $\f^{-1}(\mathbf{a})$ is the set of preimages\index{preimage set} of $\mathbf{a}$ under the block
evolution operator $\f$.
This can be generalized to the $n$-th iterate of $F$,
\begin{equation} \label{preimP}
P_n(\mathbf{a})=(F^n\mu)([\mathbf{a}])=\sum_{\mathbf{b}\in \f^{-n}(\mathbf{a})}  \mu([\mathbf{b}]),
\end{equation}
where, again,  $\f^{-n}(\mathbf{a})$ is the set of preimages of $\mathbf{a}$ under $\f^n$, the
$n$-th iterate of $\f$. Thus, if we know the elements of the set of $n$-step preimages of the block
$\mathbf{a}$ under the block evolution operator $\f$, then we can easily compute the probability
 $P_n(\mathbf{a})$.

Now, let us suppose that the initial measure is a Bernoulli measure\index{Bernoulli measure} $\mu_p$, defined by
$\mu_p([\mathbf{a}])=p^{\#_1(\mathbf{a})}(1-p)^{\#_0(\mathbf{a})}$, where $\#_s(\mathbf{a})$ denotes number of symbols $s$
in $\mathbf{a}$ and where $p\in[0,1]$ is a parameter. In such a case eq. (\ref{preimP}) reduces to 
\begin{equation} \label{bernoulliP}
P_n(\mathbf{a})=\sum_{\mathbf{b}\in \f^{-n}(\mathbf{a})}  
p^{\#_1(\mathbf{b})}(1-p)^{\#_0(\mathbf{b})}.
\end{equation}
Furthermore, if $p=1/2$, then the above reduces to even simpler form,
\begin{equation} \label{symbernoulliP}
P_n(\mathbf{a})=\sum_{\mathbf{b}\in \f^{-n}(\mathbf{a})}  
\frac{1}{2^{|\mathbf{b}|}} = \frac{\card \f^{-n}(\mathbf{a})}{2^{|\mathbf{a}|+2n}}.
\end{equation}

For many elementary CA rules and for short blocks $\mathbf{a}$, the sets $\f^{-n}(\mathbf{a})$ exhibit simple enough structure
to be described and enumerated by combinatorial methods, so that the formula for  $\card \f^{-n}$ can be
constructed and/or the sum in eq. (\ref{bernoulliP}) can be computed. Although there is no precise definition of  ``simple enough
structure'', the known cases can be informally classified into five groups:
\begin{enumerate}
 \item rules with preimage sets that are ``balanced'' (have the same number of preimages for each block),
 \item rules with preimage sets mostly composed of long blocks of identical symbols (having \emph{long runs}) or long blocks of arbitrary symbols,
 \item rules with preimage sets that can be described as sets of strings in which some local property holds everywhere,
 \item rules with preimage sets that can be described as strings in which some global (non-local) property holds,
  \item rules for which preimage sets are  related to preimage sets of some known solvable rule.
\end{enumerate}
Selection of the most interesting examples in each category is
given below. 
\subsection{Balanced preimages: surjective rules}
It is well known that the symmetric Bernoulli measure $\mu_{1/2}$ is invariant
 under the action of a surjective rule  \citep[see][and references therein]{Pivato2009}. In one dimension, surjectivity
\index{surjectivity} is a decidable property, and the relevant algorithm is known, due to  \cite{amoroso72}.
 Among elementary CA rules, 
 surjective rules\index{surjective cellular automata} have the following Wolfram numbers:  15, 30, 45, 51, 60, 90, 105, 106, 150, 154, 170 and 204. For all of them, for the initial measure
$\mu=\mu_{1/2}$ and
for any block $\mathbf{a}$, 
\begin{equation}
\label{surj}
P_n(\mathbf{a})=2^{-|\mathbf{a}|}. 
\end{equation}
The above result is a direct consequence of the Balance Theorem\index{Balance Theorem}, first proved by \cite{hedlund69},  which states that for a surjective rule, $\card \f^{-1}(\mathbf{a})$ is the same for all blocks $\mathbf{a}$ of a given length.
For elementary rules this implies that $\card \f^{-1}(\mathbf{a})=4$, and, therefore,  $\card \f^{-n}(\mathbf{a})=4^n$.
From eq. (\ref{symbernoulliP}) one then obtains eq. (\ref{surj}).

\subsection{Preimages with long runs and arbitrary symbols: rule 130}
Consider the elementary  CA with the local function
\begin{equation}
f\big(x_1, x_2, x_3\big) 
= \left\{ 
\begin{array}{l l}
  1 & \quad \text{if} \quad ( x_1 \; x_2 \; x_3) = ( 0 \; 0 \; 1) \text{ or } (1 \;1\;1),\\
 0 & \quad \text{otherwise,}\\
\end{array} \right.
\end{equation}
Its Wolfram number is $W(f)=130$, and we will refer to it as simply ``rule 130''. Subsequently,
any rule with Wolfram number $W(f)$ will be referred to as ``rule $W(f)$''.
For rule 130  and for $\mu_p$, the the probabilities  $P_n(\mathbf{a})$ are known for $|\mathbf{a}|\leq 3$ \citep{paper40}.
The corresponding formulae are rather long, thus we give only the expression for $P_n(0)$. 
\begin{equation} \label{r130P0}
P_n(0)=1-p^{2n+1}-
\frac {p\, \left( -  {p} ^{4 \lceil 
(n-2)/2 \rceil +4 }+  {p} ^{4
 \lceil (n-2)/2 \rceil +5} -  {p}^{ 4 \lfloor n/2 \rfloor +3}+{p}^{3
} - p + 1\right) }{{p}^{3}+{p}^{2}+p+1}.
\end{equation}
The above result is based on the fact that for rule 130, the set $\f^{-n}(111)$ has only one element, namely the block 
$11\ldots 1$, hence $\card \f^{-n}(111)=1$. Moreover, the set $\f^{-n}(001)$  consists of all blocks of the form
\[ \underbrace{\star \hdots \star}_{2n-2i}\; 1\; 0\; 1 \;\underbrace{1 \hdots 1}_{2i} \quad \text{(if \(i\) is odd)} \quad\quad \text{or} \quad\quad \underbrace{\star \hdots \star}_{2n-2i}\; 0\; 0\; 1 \;\underbrace{1 \hdots 1}_{2i} \quad \text{(if \(i\) is even),}\]
where $i\in \{0\dots n\}$ and \(\star\) denotes an arbitrary value in \(\mathcal{A}\). Probabilities
of occurence of blocks 111 and 001 can thus be easily computed. Using the fact that for this rule  $P_n(0) =1- P_{n-1} (111) - P_{n-1}(001)$, one then obtains eq. (\ref{r130P0}). The floor and ceiling operators appear in that formula
because different expressions are needed for odd and even $n$, as it is evident from the  structure
of preimages of 001 described above.
 Rule 130 is an example of a rule where
convergence of $P_n(0)$ to its limiting value is essentially exponential (like in rule 172 discussed below, except that
there are some small variations between values corresponding to  even and odd $n$.

 \subsection{Preimages described by a local property: rule 172}
The local function of rule 172 is defied as
\begin{equation}\label{selector}
 f(x_1,x_2,x_3) = \left \{ \begin{array}{ll}
                              x_2     & \mbox{if $x_1=0$,}\\
                              x_3     & \mbox{if $x_1=1$.}
                    \end{array}
 \right.
\end{equation}
The combinatorial structure of $\f^{-n}(\mathbf{a})$ for this rule can be described, for some blocks $\mathbf{a}$, 
as binary strings with forbidden sub-blocks. More precisely, one can prove the following proposition \citep{paper39}.
\begin{prop}
Block $\mathbf{b}$ of length $2n+1$ belongs to $\f^{-n}(1)$ for rule 172 if and only if it has the structure 
$ \mathbf{b}= \underbrace{\star \star \ldots \star}_{n-2} 001 \underbrace{\star \star \ldots \star}_{n}$,
or $\mathbf{b}= \underbrace{\star \star \ldots \star}_{n-2}  a_1 a_2\ldots a_{n+1}c_1c_2$, 
where $ a_1 a_2\ldots a_n$ is a binary string which does not contain any pair of adjacent zeros,
 and
\begin{equation} \label{condc1c2}
 c_1c_2= 
\begin{cases}
1 \star , & \text{if}\,\,\, a_{n+1}=0, \\
\star 1, & \text{otherwise}.
\end{cases}
\end{equation}
\end{prop}
 Since the number of binary strings of length $n$ without any pair of consecutive zeros is know to be  $F_{n+2}$,
where $F_n$ is the $n$-th Fibonacci number, it is not surprising that Fibonacci numbers\index{Fibonacci numbers} appear in expressions
for block probabilities of rule 172. 
 For this rule and $\mu=\mu_{1/2}$, probabilities  $P_n(\mathbf{a})$ are known for $|\mathbf{a}|\leq 3$, as shown below.
\begin{align} \label{rule172probs}
P_n(0)&=\frac{7}{8}  - \frac{F_{n+3}}{2^{n+2}},\\ \nonumber
 P_n(00) &={3}/{4}-{2}^{-n-2}F_{ n+3} -{2}^{-n-4}F_{ n+2},\\ \nonumber 
P_n(000) &={5}/{8}-{2}^{-n-2}F_{ n+3} -{2}^{-n-4}F_{ n+2},\\ \nonumber
P_n(010) &={1}/{8}-{2}^{-n-3}F_{n+1 },
\end{align}
Note that the above are probabilities in short block representation,
thus all remaining probabilities of blocks of length up to 3 can be obtained using eq. (\ref{shortform3}).
More recently,  $P_n(0)$   has been computed for arbitrary $\mu_p$ \citep{paper59}, 
\begin{equation} \label{infinitesolution}
P_n(0)=1- (1-p)^2 p
-\frac{p^2}{\lambda_2-\lambda_1} \left( \alpha_1 \lambda_1^{n-1} + \alpha_2 \lambda_2^{n-1} \right),
\end{equation}
where
\begin{align}
 \lambda_{1,2}&=\frac{1}{2}p \pm \frac{1}{2}\sqrt{p(4-3p)},\label{eigenvalues}\\
\alpha_{1,2}&=   \left( \frac{p}{2}-1 \right) \sqrt {p \left( 4-3\,p \right) } \pm  \left( \frac{p^2}{2} -1 \right).
\end{align}
\subsection{Preimages described by a non-local property: rule 184} 
While in rule 172 the ombinatorial description of sets $ \f^{-n}(\mathbf{a})$ involved some local conditions (e.g., two consecutive 
zeros are forbidden), in rule 184, with the local function
$f(x_1,x_2,x_3)=x_1 + x_2x_3 - x_1x_2$, the conditions are more of a global nature, that is, involving properties of longer substrings. 
In particular, one can show the following.
\begin{prop}
The block $b_1 b_2 \ldots b_{2n+2}$ belongs to $\f^{-n}(00)$ under rule
$184$ if and only if
$b_1=0$, $b_2=0$ and $2+\sum_{i=3}^{k} \xi(b_i) > 0$ for every
$3 \leq k \leq 2n+2$, where $\xi(0)=1$, $\xi(1)=-1$.
\end{prop}
Proof of this property relies on the fact that rule 184 is known to be equivalent to a ballistic annihilation\index{ballistic annihilation} process \citep{Krug88,paper11,Belitsky2005InvariantMA}. 
Another crucial property of rule 184 is that it is number-conserving\index{number-conserving rule}, that is, conserves the number of zeros and ones.
Using this fact and the above proposition,
probabilities $P_n(\mathbf{a})$ can be computed for for $\mu_p$ and $|\mathbf{a}|\leq 2$,
\begin{equation} 
P_n(0)=1-p, \,\,\,\,\,P_n(00) = \sum_{j=1}^{n+1} \frac{j}{n+1} {{2n+2} \choose
 {n+1-j}} p^{n+1-j} (1-p)^{n+1+j}.
\end{equation}
The main idea which is used in deriving the above expression is the fact that preimage sets $\f^{-1}(00)$ have a similar structure
 to trajectories of one-dimensional random walk starting from the origin and staying on the positive semi-axis.
Enumeration of such trajectories is a well known combinatorial problem, and the binomial coefficient appearing
in the expression for $P_n(00)$ indeed comes from this enumeration procedure. In the limit of large $n$ one can demonstrate
that
\begin{equation} \label{plimit}
 \lim_{n \rightarrow \infty} P_n(00)=
 \left\{ \begin{array}{ll}
 1-2p  & \mbox{if $p<1/2$}, \\
 0    & \mbox{otherwise}.
\end{array}
\right.
\end{equation}
All the above results  can be extended to generalizations 	of rule 184 with larger radius \citep{paper11}.

A special case of $\mu_{1/2}$ is especially interesting, as in this case probabilities of blocks up to length 3 can be  obtained,
\begin{align} 
P_n(0)&=\frac{1}{2},\\
P_n(00) &= {2}^{-2-2\,n}{2\,n+1\choose n+1},\\
P_n(000)&={2}^{-2\,n-3}{2\,n+1\choose n+1}, \\
%P_n(010)&={2}^{-3-2\,n} \left( {2}^{2n+2}-3\,{2\,n+1\choose n+1} \right) \\
P_n(010)&=\frac{1}{2}- 3 \cdot {2}^{-3-2\,n}{2\,n+1\choose n+1}.  
\end{align}
Using Stirling's approximation for factorials for large $n$, one obtains $P_n(00) \sim n^{-1/2}$,
thus  $P_n(00)$ converges to $0$ as a power law\index{power law convergence} with exponent $1/2$. 

\subsection{Preimage sets related to preimages of other solvable rules: rule 14}
The local function of rule 14 is defied by $f(0,0,1)=f(0,1,0)=f(0,1,1)=1$, and $f(x_0,x_1,x_2)=0$
for all other triples $(x_0,x_1,x_2)\in \{0,1\}^3$. 
For rule 14 and $\mu=\mu_{1/2}$, the probabilities  $P_n(\mathbf{a})$ are known for $|\mathbf{a}|\leq 3$, and are given by
\begin{align}
 P_n(0)&=\frac{1}{2}\left(1+\frac{2 n-1}{4^n}C_{n-1} \right),\\
 P_n(00)&={2}^{-2-2\,n}(n+1)C_{{n}}+\frac{1}{4},\\
 P_n(000)&= 2^{-2n-3}\left( 4\,n+3 \right) C_{n},\\
 P_n(010)&={2}^{-2-2\,n} \left( n+1 \right) C_{{n}},
\end{align}
where $C_n$ is the $n$-th Catalan number\index{Catalan numbers} \citep{paper34}.
These formulae were obtained using the fact that this rule conserves
the number of blocks 10 and that the combinatorial structure of preimage sets of some short blocks
resembles the structure of related preimage sets under the  rule 184. More precisely, computation of the above block probabilities
 relies on the following property (see  \emph{ibid.} for proof).
\begin{prop} 
For any  $n\in \NN$, the number of $n$-step preimages of 101 under the rule 14
is the same as the number of $n$-step preimages of 000 under the rule 184, that is,
\begin{equation}
\card \f^{-n}_{14}(101) = \card \f^{-n}_{184}(000),
\end{equation}
where subscripts 184 and 14 indicate block evolution 
operators
for, respectively, CA rules 184 and 14. Moreover, the bijection $M_n$ from the set $\f^{-n}_{184}(000)$
to the set $\f^{-n}_{14}(101)$ is defined by
\begin{equation}
M_n(x_0x_1\ldots x_m)=\left\{n+j+1+\sum_{i=0}^j x_i\mod 2\right\}_{j=0}^{m}
\end{equation}
for  $m\in \NN$ and for $x_0x_1\ldots x_m \in \{0,1\}^m$.
\end{prop}

As in the case of rule 184, one can show that for rule 14 and large $n$,
\begin{equation}
 P_n(0) \approx \frac{1}{2} +\frac{1}{4 \sqrt{\pi}}n^{-\frac{1}{2}}.
\end{equation}
The power law which appears here exhibits the same exponent  as in the case of  rule 184 for $P_n(00)$.

\subsection{Convergence of block probabilities}
The examples shown in the previous sections indicate that in all cases for which $P_n(\mathbf{a})$ can be computed exactly,
 as $n\to \infty$, $P_n(\mathbf{a})$ remains either constant, or converges to its limiting value exponentially or as a power law.
The exponential convergence is the most prevalent. Indeed, for many other elementary CA rules for which formulae for $P_n(0)$ are either known
or conjectured, the  exponential convergence\index{exponential convergence} to $P_{\infty}(0)$ can be observed most frequently.  This includes 15
elementary rules which are known as asymptotic emulators of identity \citep{Rogers94,paper52}. Formulae 
for $P_n(1)$ for the initial measure $\mu_{1/2}$ for these rules are shown below. Starred rules are those for which
a formal proof has been published in the literature \citep[see][and references therein]{paper52}.
\begin{itemize}
\item
Rule $13$:   $P_n(1)=7/16-(-2)^{-n-3}$       
\item 
Rule $32^{\star} $:   $P_n(1)=2^{-1-2n}$     
\item 
Rule $40$:   $P_n(1)=2^{-n-1}$       
\item 
Rule $44$:   $P_n(1)=1/6+\frac{5}{6}2^{-2n}$       
\item 
Rule $77^{\star}$:   $P_n(1)=1/2$       
\item 
Rule $78$:   $P_n(1)=9/16$       
\item 
Rule $128^{\star}$:   $P_n(1)=2^{-1-2n}$       
\item 
Rule $132^{\star}$:   $P_n(1)=1/6+\frac{1}{3}2^{-2n}$       
\item 
Rule $136^{\star}$:   $P_n(1)=2^{-n-1}$       
\item 
Rule $140^{\star}$:   $P_n(1)=1/4+2^{-n-2}$         
\item 
Rule $160^{\star}$:   $P_n(1)=2^{-n-1}$       
\item 
Rule $164$:   $P_n(1)=1/12-\frac{1}{3}4^{-n}+\frac{3}{4}2^{-n}$       
\item 
Rule $168^{\star}$:   $P_n(1)=3^{n}2^{-2n-1}$       
\item 
Rule $172^{\star}$:   $P_n(1)=\frac{1}{8}+\frac{(10-4\sqrt{5})(1-\sqrt{5})^{n}+(10+4\sqrt{5})(1+\sqrt{5})^{n}}{40\cdot2^{2n}}$       
\item 
Rule $232$:   $P_n(1)=    1/2$       
\end{itemize}
The formula for rule 172, included here for completeness, can obviously be obtained from eq. (\ref{rule172probs}) by using explicit 
expressions for Fibonacci numbers in terms of powers of the golden ratio\index{golden ratio}.

Power laws appearing in rules 184 and 14, as mentioned already, are a result of the fact that dynamics of these rules
can be understood as a motion of deterministic ``particles'' propagating in a regular background. The same type of ``defect kinematics''\index{defect kinematics} has been observed, among other elementary CA rules, in 
rule 18 \citep{GRASSBERGER198452}, for which
$$P_n(11) \sim n^{-1/2}.$$
The above power law can be explained by the fact that in rule 18 one can view sequences of 0's of even length as ``defects''
which  perform a random walk\index{random walk} and annihilate upon collision, as discovered numerically
 by \cite{GRASSBERGER198452} and later formally demonstrated by \cite{Eloranta1992}. A very general treatment of 
particle kinematics in CA confirming this result can be found in  the work of \cite{PIVATO2007205}. 

Another example of an interesting power law appears in  rule 54, for which \cite{BNR91}  numerically verified that
 $$P_n(1) \sim n^{-\gamma},$$
where $\gamma \approx 0.15$. Particle kinematics of rule 54 is now very well understood \citep{PIVATO2007205}, but the above power law  has not been 
formally demonstrated, and  the exact value of the exponent $\gamma$ remains unknown.

\section{Examples of exact results for probabilistic  CA rules}
For probabilistic rules, one cannot use eq. (\ref{bernoulliP}) because the block evolution operator $\mathbf{f}^{-n}$ does
not have any obvious non-deterministic version.  Once thus has to work directly with eq. (\ref{rulefed}).

Equation (\ref{rulefed}) can be written for the $n$-th iterate of  $F$,
\begin{equation} \label{ruleiterate-nth}
(F^n\mu)([\mathbf{a}]_i)=\sum_{\mathbf{b}\in \mathcal{A}^{|\mathbf{a}|+2nr}} w^n(\mathbf{a}| \mathbf{b}) \mu([\mathbf{b}]_{i-nr})
\mathrm{\,\, for\,\, all\,\,}  i \in \mathbb{Z}, \mathbf{a} \in \mathcal{A}^{\star},
\end{equation}
where we define the {\em $n$-step block transition probability} $w^n$
recursively, so that, when $n \geq 2$ and for any blocks $c\mathbf{a} \in {\mathcal{A}}^{\star}$ and $\mathbf{b} \in {\mathcal{A}}^{|\mathbf{a}|+2rn}$, 
\begin{equation}
w^n(\mathbf{a} | \mathbf{b}) = \sum_{ \mathbf{b}' \in {\mathcal{A}}^{|\mathbf{a}|+2r(n-1)}} w^{n-1}( \mathbf{a}  | \mathbf{b}')
 w(\mathbf{b}'  | \mathbf{b}).
\end{equation}
The $n$-step block
transition probability $w^n(\mathbf{a} | \mathbf{b})$ can be intuitively understood as
 the conditional probability of seeing the block 
$\mathbf{a}$ after
$n$ iterations of $F$, conditioned on the fact that the original
configuration contained the block $\mathbf{b}$.

Using definition of $w$ given in eq. (\ref{defw}), one can produce an explicit formula for $w^n$,
\begin{equation} \label{wexplicit}
w^n(\mathbf{a} | \mathbf{b}) = \sum_{\substack{\mathbf{b}_{n-1} \in {\mathcal{A}}^{|\mathbf{a}|+2r(n-1)} \vspace{-1.5mm}\\
\vdots \vspace{0.5mm}\\
\mathbf{b}_1 \in \mathcal{A}^{|\mathbf{a}|+2r}}} 
w(\mathbf{a} | \mathbf{b}_1) \left( \prod_{i=1}^{n-2}
w(\mathbf{b}_i | \mathbf{b}_{i+1}) \right) w(\mathbf{b}_{n-1} | \mathbf{b}).
\end{equation}
For a shift-invariant initial  probability measure $\mu$, equation (\ref{ruleiterate-nth}) becomes
\begin{equation} \label{ruleiterate-nth-new}
P_n(\mathbf{a})=\sum_{\mathbf{b}\in \mathcal{A}^{|\mathbf{a}|+2nr}} w^n(\mathbf{a}| \mathbf{b}) P_0(\mathbf{b}).
\end{equation}
Since some of the transition probabilities may be zero, we define, 
for any block $\mathbf{b} \in \mathcal{A}^{\star}$,  
\begin{equation}
\supp  w^n(\mathbf{a}| \cdot)=\{  \mathbf{b} \in \mathcal{A}^{|\mathbf{a}+2nr}: w^n(\mathbf{a}|\mathbf{b}) > 0\},
\end{equation}
and then we have
\begin{equation} \label{probrulemethod}
P_n(\mathbf{a})=\sum_{\mathbf{b}\in \supp  w^n(\mathbf{a}| \cdot)} w^n(\mathbf{a}| \mathbf{b}) P_0(\mathbf{b}).
\end{equation}
 In some cases, $\supp  w^n(\mathbf{a}| \cdot)$ is small and has a simple structure, and the needed $ w^n(\mathbf{a}| \mathbf{b})$
can be computed directly from eq. (\ref{wexplicit}).
This approach has been successfully used for a class of probabilistic CA rules known as $\alpha$-asynchronous rules with single transitions \citep{paper44}. We show two examples of such rules below.
\subsection{Rule 200A}
 Rule 200A, known as $\alpha$-asynchronous rule 200, is defined by transition probabilities
\begin{equation}
 w(1 | \mathbf{b}) =  \begin{cases} 0 & \quad \text{if\;} \mathbf{b} \in \{000,001,100,101\},
\\
1 & \quad \text{if\;} \mathbf{b} \in \{011,110,111\}, \\
1-\alpha & \quad \text{if\;} \mathbf{b} =010,  \end{cases}
\end{equation}
and $w(0|\mathbf{b})=1-w(1|\mathbf{b})$ for all $\mathbf{b} \in \{0,1\}^3$, where $\alpha \in [0,1]$ is a parameter. 
The set $\supp w^n(1| \cdot)$ for this rule consists of all blocks of the form
\begin{equation}
\underbrace{\star \; \cdots \; \star\;}_{n}\; 1 \;\underbrace{ \star \;
\cdots \; \star}_{n}.
\end{equation}
Moreover, one can show that for any block $\mathbf{b} \in \supp w^n(1| \cdot)$ where the central 1 has 0s as both neighbours,
$w^n(1|\mathbf{b}) = (1-\alpha)^n$, while $w^n(1|\mathbf{b}) = 1$ for all other cases. This allows to compute
$P_n(1)$ and $P_n(0)$ from eq. (\ref{probrulemethod}) directly. By the same method  probabilities of other blocks
of length up to 3 can be computed for rule 200A and $\mu_p$, and results are shown below.
 \begin{align}
 P_n(0)&=1-{p}^{2} \left( 2-p \right) - \left( 1-p \right) ^{n}p\,
  \left( 1-p \right) ^{2},\\ \nonumber
%                                    "Ala"
P_n(00)&=- \left( -1+p \right) ^{2} \left( -2\,p+2\, \left( 1-\alpha
 \right) ^{n}p-1 \right), \\ \nonumber
 %%                                   "Ala"
P_n(000)&=-{p}^{2} \left( -1+p \right) ^{3} \left( 1-\alpha \right) ^{2\,n
}+p\, \left( 2\,{p}^{2}-3 \right)  \left( -1+p \right) ^{2}
 \left( 1-\alpha \right) ^{n}\\ \nonumber
 &- \left( {p}^{3}+{p}^{2}-2\,p-1
 \right)  \left( -1+p \right) ^{2},\\ \nonumber
 %%                                  "Ala"
 P_n(010)&=\left( 1-\alpha \right) ^{n}p\, \left( 1-p \right) ^{2}.
\end{align}
Exponential convergence toward limiting values can clearly be observed in all of these block probabilities.
\subsection{Rule 140A}
Another example is rule 140A, defined as
\begin{equation}
w(1 | \mathbf{b}) =  \begin{cases} 0 & \quad \text{if\;} \mathbf{b} \in \{000,001,100,101\},
\\
1 & \quad \text{if\;} \mathbf{b} \in \{010,011,111\}, \\
1-\alpha & \quad \text{if\;} \mathbf{b} = 110,  \end{cases}
\end{equation}
and $w(0|\mathbf{b})=1-w(1|\mathbf{b})$ for all $\mathbf{b} \in \{0,1\}^3$. For this rule the  set $\supp w^n(1| \cdot)$
has the same structure as for rule 200A, except that the values of $w^n(1|\mathbf{b})$ are different.
For blocks $\mathbf{b}\in \mathcal{A}^{2n+1}$ having the structure
\begin{equation}  \underbrace{\star \; \cdots \; \star\;}_{n-1}1 \quad\underline{1} \quad
\underbrace{ 1 \; \cdots \; 1 }_{k-1}\; 0 \;\underbrace{ \star \; \cdots \;
\star}_{n-k} ,
\end{equation}
where $0 \leq k-1 \leq n$, it has been demonstrated  \citep{paper44} that
\begin{equation}
w^n(1|\mathbf{b}) = \begin{cases} \beta^n &\quad \text{if } k=1, \\
 \beta^n \left(\frac{\alpha}{\beta}\right)^{k-1} \binom{n-1}{k-1} +
\beta^{n-k+1} \sum_{j=0}^{k-2}\limits \binom{n-k+j}{j} \alpha^j 
 &\quad \text{if } 2 \leq k \leq n, \\
1 &\quad \text{if } k = n+1. \end{cases}
\end{equation}
For all other blocks in $\supp w^n(1| \cdot)$ one has $w^n(1|\mathbf{b})=1$.
Using this result, probability $P_n(0)$ can be computed assuming initial measure $\mu_p$, although the
summation in eq. (\ref{ruleiterate-nth}) is rather complicated. The end result, shown below, is nevertheless
surprisingly simple.
\begin{equation}
 P_n(0) =  1-\rho(1-\rho)  - \rho^2\left( 1-(1-\rho)\alpha\right)^{n}.
\end{equation}
Corresponding formulae for  $P_n(\mathbf{a})$  for all $|\mathbf{a}|\leq 3$ have been constructed as well, but
are omitted here.

\subsection{Complete sets}
Another case when  eq. (\ref{recP}) becomes solvable is when there exists a subset of blocks
 which is called \emph{complete}\index{complete set}. A set of words $\mathcal{A}^{\star} \supset C=\{\mathbf{a}_1,\mathbf{a}_2,\mathbf{a}_3,\ldots\}$ is
\emph{complete}
with respect to a CA rule $F$ if for every  $\mathbf{a}\in C$ and $n\in\mathbb{N}$,  $P_{n+1}(\mathbf{a})$ can be expressed
as a linear combination of $P_n(\mathbf{a}_1),P_n(\mathbf{a}_2),P_n(\mathbf{a}_3),\ldots$. In this case, one can write
eqs. (\ref{recP}) for blocks of the complete set only, and the right hand sides of them will also only include probabilities
of blocks from the complete set. This way, a well-posed system of recurrence equations is obtained, and (at least in principle) it
should be solvable.

This approach has been recently applied to a probabilistic CA rule defined by
\begin{align} \label{adpodef}
 w(1|000)&=0,\, w(1|001)=\alpha,\,w(1|010)=1,\,w(1|011)=1,\\
 w(1|100)&=\beta,\,w(1|101)=\gamma,\,w(1|110)=1,\,w(1|111)=1, \nonumber
\end{align}
and $w(0|b)=1-w(1|b)$ for all $b \in \{0,1\}^3$, where  $\alpha,\beta,\gamma \in[0,1]$ are fixed parameters. This rule can be viewed as a generalized simple model
for diffusion of innovations on one-dimensional lattice \citep{paper54}. The complete set for this rule consists of
blocks $101$, $1001$, $100001$, $\ldots$. Equations (\ref{recP}) for blocks of the complete set simplify to
\begin{equation} \label{P101rec}
 P_{n+1}(101)=(1-\gamma)P_n(101)+(\alpha-2\alpha \beta+\beta)P_n(1001)+\alpha \beta P_n(10001),
\end{equation}
and, for $k>1$, 
\begin{equation} \label{P10n1rec}
 P_{n+1}(10^k1)=(1-\alpha)(1-\beta)P_n(10^k1)+(\alpha-2\alpha \beta+ \beta)P_n(10^{k+1}1)+\alpha \beta P_n(10^{k+2}1).
\end{equation}
The above equations can be solved, and, using the cluster expansion formula \citep{Stauffer94},
\begin{equation} \label{clusterexpansion}
 P_n(0)=\sum_{k=1}^\infty k P_n(10^k1),
\end{equation}
one obtains, assuming that  the initial measure is $\mu_p$,
\begin{equation}
P_n(0)= \begin{cases}
   E \left( \left( {p}\,\beta -1 \right)  \left( {p}\, \alpha-1 \right) 
 \right) ^{n} +F \left( 1-\gamma \right) ^{n} & \text{if } \alpha \beta p^2-(\alpha+\beta) p+\gamma \neq 0, \\
   (G+Hn)(1-\gamma)^{n-1}      & \text{if } \alpha \beta p^2-(\alpha+\beta) p+\gamma=0,
  \end{cases}
\end{equation}
where $E,F,G,H$ are constants depending on parameters $\alpha, \beta, \gamma$ and $p$ \citep[for detailed formulae, see][]{paper54}. For $\alpha \beta p^2-(\alpha+\beta) p+\gamma=0$, this is an example of a linear-exponential convergence of $P_n(0)$ toward its limiting value,
the only one known for a binary rule. 

\section{Future directions}
Both approximate and exact methods for computing orbits of Bernoulli measures under the action of cellular automata
need further development.

Regarding approximate methods,  although some simple classes of CA rules for which local structure approximation becomes exact are known, it is not 
known if there exist any wider classes of non-trivial rules for which this would be the case. This is certainly an area which
needs further research. There seems to be some evidence that orbits of  many deterministic rules possessing additive invariants 
are very well approximated by local structure theory, but no general results are known.

Regarding exact methods, the situation is similar. Although methods for computing exact values of $P_n(\mathbf{a})$ presented here are applicable to many different rules, it is still
not clear if they are applicable to some  wider classes of CA in general. Some such classes has been proposed, but formal results are still lacking. For example, there is a number of rules for which 
convergence of $P_n(1)$ to its limiting value $P_{\infty}(1)$ is known to be exponential, and it has been conjectured that
for all rules known as \emph{asymptotic emulators of identity}\index{asymptotic emulators of identity} this is indeed the case. However, there
seems to be some recent evidence that for rule 164, which belongs to asymptotic emulators of identity, the convergence is
not exactly exponential  (A. Skelton, private communication). Are then other classes of CA rules for which the convergence is always exponential?
And, more importantly, are there any wide classes of non-trivial CA for which exact formulae for probabilities
of short block are obtainable?

Another interesting question is the relationship between exact orbits of CA rules and approximate orbits obtained
by iterating local structure maps. Which features or exact orbits are ``inherited'' by approximate orbits?
It seems that often existence of additive invariants is ``inherited'' by local structure maps, yet more work in this direction
is needed. On a related note, such behavior of $P_n(1)$ as observed in rules 172 or 140A (discussed earlier in this article) strongly resembles hyperbolicity 
in finitely-dimensional dynamical systems.  Hyperbolic fixed points are common type of fixed points in dynamical systems.
If the initial value is near the fixed point and lies on the stable manifold, 
the orbit of the dynamical system converges to the fixed point exponentially fast.
One could argue that the exponential convergence to $P_{\infty}(1)$ observed
in such  rules as rule 172 or 140A is somewhat related to finitely-dimensional hyperbolicity.
Since local structure maps which approximate dynamics of a given CA are finitely-dimensional, 
one could ask what is the nature of their fixed points -- are these hyperbolic for CA exhibiting
hyperbolic-like dynamics?  Is hyperbolicity of orbits of CA rules somewhat ``inherited'' by local structure maps?
If so, under what conditions does this happen? All those questions need to be investigated in details in future years.

Finally, one should mention that both theoretical developments and examples presented in this article 
pertain to one-dimensional cellular automata. Higher-dimensional systems have been studied in the context
of the local structure theory \citep{gutowitz87b}, and some examples or two-dimensional cellular automata
with exact expressions for small block probabilities are known \citep{paper41}, yet the
 orbits of Bernoulli measures under higher-dimensional CA are still mostly an unexplored terrain. 
Given the importance of two- and threepice
-dimensional CA in applications, this subject
will likely attract some attention in the near future.

\printindex
\end{document}